\documentclass[11 pt,a4paper,usenames,dvipsnames]{article}
\usepackage[utf8]{inputenc}
\usepackage{amsmath}
\usepackage{amsfonts}
\usepackage{amssymb}
\usepackage{pgfplots}
\usepackage{xcolor}

\usepgfplotslibrary{polar}
\usepackage{tikz}
\usepackage[english]{babel}
\newtheorem{theorem}{Theorem}

\usepackage{subcaption}
\usepgfplotslibrary{fillbetween}
\usetikzlibrary{patterns}
\usetikzlibrary{decorations.pathreplacing}
\usepackage{indentfirst}
\usepackage{amsmath}

\author{Dr. Jacques Balayla MD, MPH, CIP, FRCSC\footnote{To whom correspondence should be addressed: Dr. Jacques Balayla MD, MPH, CIP, FRCSC. Quilligan Scholar. e-mail: jacques.balayla@mcgill.ca. Osler Fellow. Department of Obstetrics and Gynecology. McGill University, Montreal, Quebec, Canada}}

\title{Theorems on the Geometric Definition \\ of the Positive Likelihood Ratio (LR+)}

\date{}

\begin{document}
\maketitle  
\begin{abstract}
From the fundamental theorem of screening (FTS) we obtain the following mathematical relationship relaying the pre-test probability of disease $\phi$ to the positive predictive value $\rho(\phi)$ of a screening test:
\begin{center}
\begin{large}
$\displaystyle\lim_{\varepsilon \to 2}{\displaystyle \int_{0}^{1}}{\rho(\phi)d\phi} = 1$
\end{large}
\end{center}
\

where $\varepsilon$ is the screening coefficient - the sum of the sensitivity ($a$) and specificity ($b$) parameters of the test in question. However, given the invariant points on the screening plane, identical values of $\varepsilon$ may yield different shapes of the screening curve since $\varepsilon$ does not respect traditional commutative properties. 
In order to compare the performance between two screening curves with identical $\varepsilon$ values, we derive two geometric definitions of the positive likelihood ratio (LR+), defined as the likelihood of a positive test result in patients with the disease divided by the likelihood of a positive test result in patients without the disease, which helps distinguish the performance of both screening tests. The first definition uses the angle $\beta$ created on the vertical axis by the line between the origin invariant and the prevalence threshold $\phi_e$ such that $LR+ = \frac{a}{1-b} = cot^2{(\beta)}$.  The second definition projects two lines $(y_1,y_2)$ from any point on the curve to the invariant points on the plane and defines the LR+ as the ratio of its derivatives  $\frac{dy_1}{dx}$ and $\frac{dy_2}{dx}$. Using the concepts of the prevalence threshold and the invariant points on the screening plane, the work herein presented provides a new geometric definition of the positive likelihood ratio (LR+) throughout the prevalence spectrum and describes a formal measure to compare the performance of two screening tests whose screening coefficients $\varepsilon$ are equal.
\end{abstract}
\newpage
\section{The Fundamental Theorem of Screening}
From the fundamental theorem of screening we obtain the following mathematical relationship relaying the positive predictive value ($\rho$) to the pre-test probability ($\phi$), which equals the prevalence of disease amongtst individuals at baseline risk \cite{manrai2014medicine}:
\begin{large}
\begin{equation}
\displaystyle\lim_{\varepsilon \to 2}{\displaystyle \int_{0}^{1}}{\rho(\phi)d\phi} = 1
\end{equation}
\end{large}
\

where $\varepsilon$ is equal to the sum of the sensitivity ($a$) and specificity ($b$) parameters of a screening test in question \cite{balayla2020prevalence}. Equation (1) holds since the Euclidean plane, henceforth referred to as the screening plane, which contains the domain and range of the screening curve is a square of dimensions 1 x 1 and consequently, of area 1.
\subsection{The Screening Plane}
Graphically, the screening plane can be depicted as follows, with the vertical axis representing the positive predictive value and the horizontal axis representing the pre-test probability or prevalence of disease:
\begin{center}

\begin{tikzpicture}
 
	\begin{axis}[
    axis lines = left,
    xlabel = $\phi$,
	ylabel = {$\rho(\phi)$},    
     ymin=0, ymax=1,
    legend pos = south east,
     ymajorgrids=true,
     xmajorgrids=true,
    grid style=dashed,
    width=6cm,
    height=6cm,
     ]
	\end{axis} 
\end{tikzpicture}
\\
\
\textbf{Figure 1. The Screening Plane}
\end{center}
\

As can be readily observed, the range and domain of the screening curve span from [0-1], and all curves share $at$ $least$ two invariant points at [0,0] and [1,1] \cite{balayla2020prevalence}. While we're restricting the screening plane to this square in order to obtain clinically useful information, the screening curve extends beyond this area.
\newpage
\section{Invariant Points}
In mathematics, a fixed point (sometimes shortened to fixpoint, also known as an invariant point) of a function is an element of the function's domain that is mapped to itself by the function. That is to say, c is a fixed point of the function $f$ if $f$(c) = c \cite{birkhoff1922invariant}. In other words, there is a point with coordinates $\lbrace c, f(c)\rbrace$ that equals $\lbrace c,c\rbrace$. Invariant points do not move when a specific transformation is applied. However, points which are invariant under one transformation may not be invariant under a different transformation \cite{birkhoff1922invariant}. 
We can illustrate graphically the invariant points at the extremes of the screening plane at [0,0] and [1,1], with multiple screening curves with different sensitivity and specificity parameters in the same plane as follows:
\begin{center}

\begin{tikzpicture}
 
	\begin{axis}[
    axis lines = left,
    xlabel = $\phi$,
	ylabel = {$\rho(\phi)$},    
     ymin=0, ymax=1,
    legend pos = south east,
     ymajorgrids=true,
     xmajorgrids=true,
    grid style=dashed,
    width=6cm,
    height=6cm,
     ]
     \addplot [
	domain= 0:1,
	color= blue,
	]
	{(0.95*x)/((0.95*x+(1-0.99)*(1-x))};
	 \addplot [
	domain= 0:1,
	color= orange,
	]
	{(0.85*x)/((0.85*x+(1-0.95)*(1-x))};
	\addplot [
	domain= 0:1,
	color= red,
	]
	{(0.75*x)/((0.75*x+(1-0.85)*(1-x))}; 
	\addplot [
	domain= 0:1,
	color= gray,
	]
	{(0.5*x)/((0.5*x+(1-0.5)*(1-x))};
	\addplot [
	domain= 0:1,
	color= black,
	]
	{(0.2*x)/((0.2*x+(1-0.4)*(1-x))};
	\addplot [
	domain= 0:1,
	color= magenta,
	]
	{(0.1*x)/((0.1*x+(1-0.1)*(1-x))};
	\addplot [
	domain= 0:1,
	color= brown	,
	]
	{(0.02*x)/((0.02*x+(1-0.02)*(1-x))};
	\end{axis}
\end{tikzpicture}

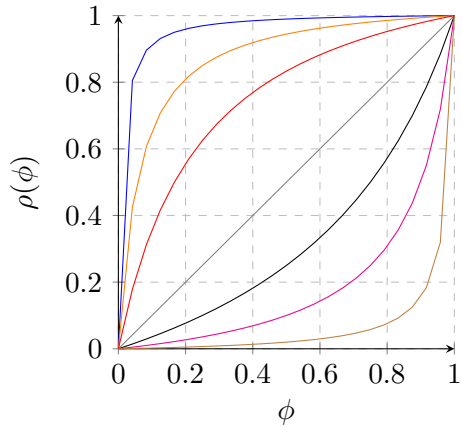
\textbf{Figure 2. Screening curves with different $\varepsilon$ values}

\end{center}

\section{Non-Commutative Properties of $\varepsilon$}
Two identical values of $\varepsilon$ will yield different screening curves depending on the individual values of the sensitivity and specificity. In this sense, the make-up of $\varepsilon$ is non-commutative because the equation for the positive predictive value isn't linear. Take as an example two different tests whose $\varepsilon$ value equal to 1.70. The first case we have a sensitivity of 95$\%$ and a specificity of 75$\%$. In the second case, we have the reverse, a sensitivity of 75$\%$ and a specificity of 95$\%$ \cite{balayla2020derivation}. 
\newpage
Graphically, these scenarios yield the following curves:
\begin{center}
\begin{tikzpicture}
 
	\begin{axis}[
    axis lines = left,
    xlabel = $\phi$,
	ylabel = {$\rho(\phi)$},    
     ymin=0, ymax=1,
    legend pos = south east,
     ymajorgrids=true,
     xmajorgrids=true,
    grid style=dashed,
    width=6cm,
    height=6cm,
     ]
     \addplot [
	domain= 0:1,
	color= blue,
	]
	{(0.95*x)/((0.95*x+(1-0.75)*(1-x))};
	 \addplot [
	domain= 0:1,
	color= orange,
	]
	{(0.75*x)/((0.75*x+(1-0.95)*(1-x))};
	
	\end{axis}
\end{tikzpicture}

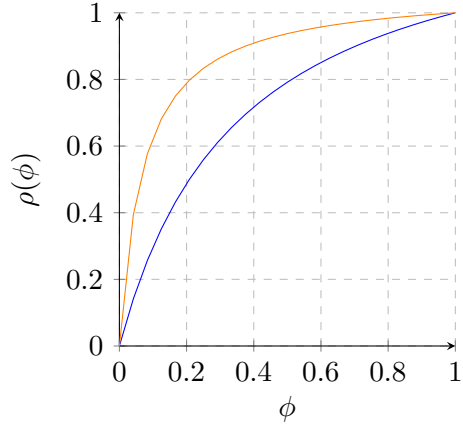
\textbf{Figure 3. Identical values of $\varepsilon$ yield different screening curves}
\end{center}
It is clear therefore that despite $\varepsilon_1$ = $\varepsilon_2$ = 1.70, the areas under the curve are different such that:
\

\begin{center}
\begin{large}
${\displaystyle \int_{0}^{1}}{{\rho_2}(\phi)d\phi}>{\displaystyle \int_{0}^{1}}{{\rho_1}(\phi)d\phi}$
\end{large}
\end{center}

To overcome the non-commutative properties of $\varepsilon$ and determine which tests performs better under standard conditions, we can take advantage of the invariant points of the screening curve at the extremes of the domain, namely the origin [0,0] and the endpoint [1,1].

\begin{center}

\begin{tikzpicture}
 
	\begin{axis}[
    axis lines = left,
    xlabel = $\phi$,
	ylabel = {$\rho(\phi)$},    
     ymin=0, ymax=1,
     xmin=0,xmax=1,
    legend pos = south east,
     ymajorgrids=true,
     xmajorgrids=true,
    grid style=dashed,
    width=6cm,
    height=6cm,
     ]
     \addplot[mark=*,color=red] coordinates {(0,0)};
     \addplot[mark=*,color=red] coordinates {(1,1)};
	\end{axis} 
\end{tikzpicture}
\\
\
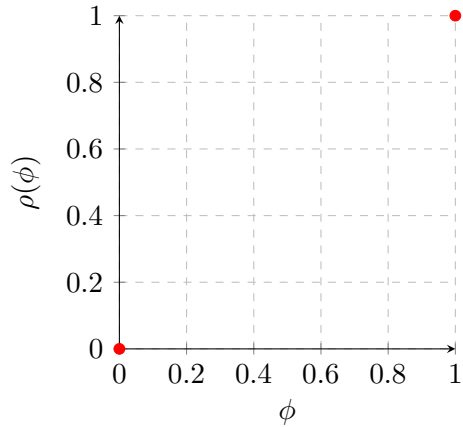
\textbf{Figure 4. Invariant points (red) on the screening plane}
\end{center}
\

\section{The Likelihood Ratio}
The likelihood ratio is a computable statistic which provides a direct estimate of how much a test result will change the odds of having a disease \cite{hayden1999likelihood}. In essence, the likelihood ratio (LR) for a dichotomous test is defined as the likelihood of a test result in patients with the disease divided by the likelihood of the test result in patients without the disease. Both the positive and negative likelihood ratios can be calculated depending on the clinical scenario by simply using the sensitivity and specificity parameters of a test. A LR close to 1 means that the test result does not change the likelihood of disease or the outcome of interest appreciably \cite{hayden1999likelihood}. The more the likelihood ratio for a positive test (LR+) is greater than 1, the more likely the disease or outcome \cite{grove1984positive}. The more a likelihood ratio for a negative test is less than 1, the less likely the disease or outcome. Thus, LRs correspond to the clinical concepts of ruling in and ruling out disease \cite{hayden1999likelihood}. 
\section{The Prevalence Threshold}
We have previously defined the prevalence threshold as the prevalence level on the screening curve below which  screening tests start to produce an increasing amount of false positive results \cite{balayla2020prevalence}. In technical terms, this is equivalent to the inflection point, also known as  point of greatest curvature, on the screening curve below which the the rate of change of a test's positive predictive value drops at a differential pace relative to the prevalence \cite{balayla2020prevalence}. This value, termed $\phi_e$, is defined at the following point on the prevalence (pre-test probability) axis:
\
\begin{large}
\begin{equation}
\phi_e = \frac{\sqrt{a\left(-b+1\right)}+b-1}{(\varepsilon-1)}=\frac{\sqrt{1-b}}{\sqrt{a}+\sqrt{1-b}}
\end{equation}
\end{large}
\begin{theorem}
Let S be the screening plane of area 1 with invariant points [0,0] and [1,1] where lies the screening curve's continuous function $0<\rho(\phi)<1$. Then there is a vertical function $f(a,b)$ which transects the screening curve at the prevalence threshold, such that the square of the cotangent of the angle $\beta$ formed between the line $y_1$ from the origin to the prevalence threshold and the vertical axis equals the positive likelihood ratio (LR+).
\end{theorem}
\subsection{Proof of Theorem 1}

If we take the point [0,0], we can draw a line from the origin to the intersection of the prevalence threshold with the screening curve. The coordinates of this intersecting point are [$\frac{\sqrt{1-b}}{\sqrt{a}+\sqrt{1-b}}$, $\sqrt{\frac{a}{1-b}}\frac{\sqrt{1-b}}{\sqrt{a}+\sqrt{1-b}}$]. This intersection allows us to determine the slope $m$ of the line that crosses that point and the origin at [0,0] (Figure 5).

\begin{large}
\begin{equation}
m = \frac{\Delta y}{\Delta x} = \frac{y_2-y_1}{x_2-x_1}=\frac{\sqrt{\frac{a}{1-b}}\frac{\sqrt{1-b}}{\sqrt{a}+\sqrt{1-b}}-0}{\frac{\sqrt{1-b}}{\sqrt{a}+\sqrt{1-b}}-0} = \sqrt{\frac{a}{1-b}}
\end{equation}
\end{large}

Since by definition the point [0,0] falls on the line, the linear equation from the origin to the prevalence  threshold point is therefore simply:

\begin{large}
\begin{equation}
f(x) = \sqrt{\frac{a}{1-b}}x
\end{equation}
\end{large}
\

where $\frac{a}{1-b}$ is the positive likelihood ratio (LR+), defined as the likelihood of a positive test result in patients $with$ the disease divided by the likelihood of a positive test result in patients $without$ the disease. To distinguish between two screening curves with identical $\varepsilon$ values, we make use of the angle $\beta$ created by the line between the invariant origin and the prevalence threshold $\phi_e$ to make a right-angle triangle as follows:

\begin{center}
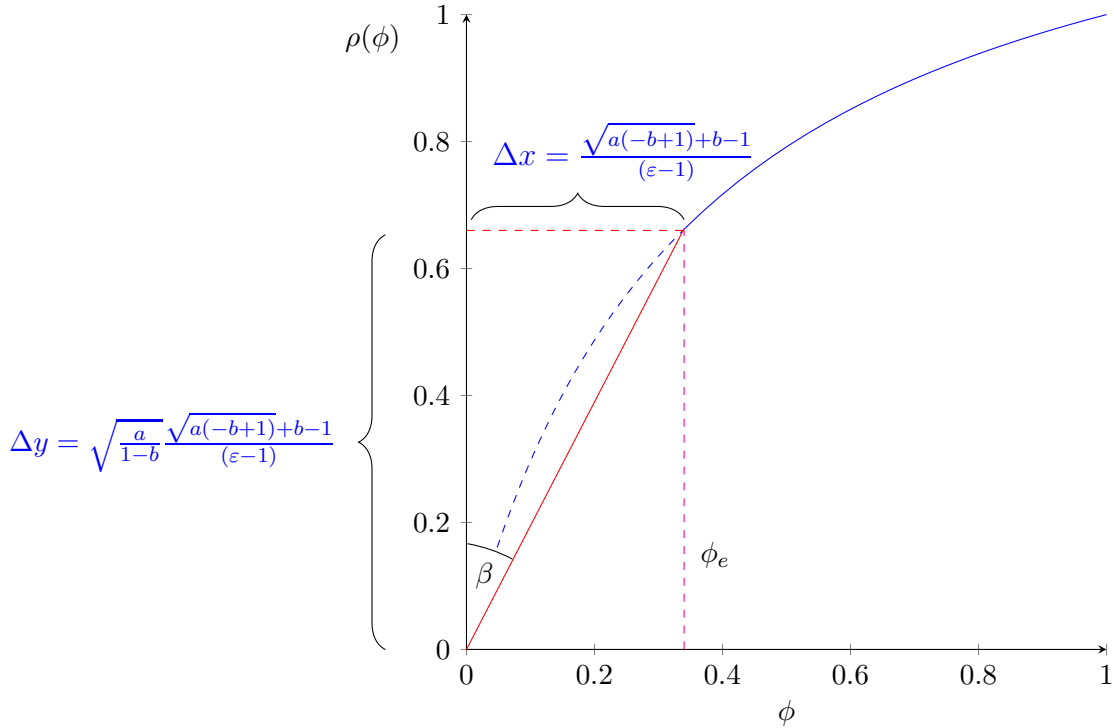
\textbf{Figure 5. Representation of the angle $\beta$ on the screening curve}
\\
\

\hspace*{-3.7em}
\begin{tikzpicture}
 
	\begin{axis}[
    axis lines = left,
    xlabel = $\phi$,
	ylabel = {$\rho(\phi)$},
	   y label style={at={(current axis.above origin)},rotate=270,anchor=north},    
     ymin=0, ymax=1,
     xmin=0,xmax=1,
    legend pos = south east,
    width=10cm,
    height=10cm,
     ]
      \addplot [
	domain= 0.34:1,
	color= blue,
	]
	{(0.95*x)/((0.95*x+(1-0.75)*(1-x))};
	 \addplot [
	domain= 0.048:0.34,
	color= blue,
	style=dashed,
	]
	{(0.95*x)/((0.95*x+(1-0.75)*(1-x))};
	 \addplot [
	domain= 0:0.34,
	color= red,
	]
	{((0.95/(1-0.75))^0.5)*x};
	 \addplot [
	domain= 0:0.34,
	color= red,
	style=dashed,
	]
	{0.66};
	\addplot[samples=50, smooth,domain=0:6,magenta, name path=three, style=dashed] coordinates {(0.34,0)(0.34,0.66)};
	\end{axis}
    \draw (0.6,1.2) arc (62:80:2);
    \node[] at (76.5:1.0) {$\beta$};
    \node[] at (21:3.5)  {$\phi_e$};
	\draw [decorate,decoration={brace,amplitude=10pt},xshift=-2pt,yshift=0pt]
(-1,0) -- (-1,5.5) node [blue,midway,xshift=-2.8cm] 
{$\Delta y = \sqrt{\frac{a}{1-b}}\frac{\sqrt{a\left(-b+1\right)}+b-1}{(\varepsilon-1)}$};
\draw [decorate,decoration={brace,amplitude=10pt},xshift=-4pt,yshift=0pt]
(0.2,5.7) -- (3,5.7) node [blue,midway,xshift=0.6cm,yshift=0.9cm] 
{\large $\Delta x = \frac{\sqrt{a\left(-b+1\right)}+b-1}{(\varepsilon-1)} $};
\end{tikzpicture}

\end{center}
\
\
\newpage
We can determine the value of $\beta$ through the following trigonometric identities:

\begin{large}
\begin{equation}
tan(\beta) =\frac{opp}{adj}= \frac{\Delta x}{\Delta y} = \frac{1}{\sqrt{\frac{a}{1-b}}}=\sqrt{\frac{1-b}{a}}
\end{equation}
\end{large}
\

We can therefore isolate $\beta$ such that:
\begin{large}
\begin{equation}
\beta = arctan\left(\sqrt{\frac{1-b}{a}}\right) 
\end{equation}
\end{large}

For simplicity's sake we can define the expression $\sqrt{\frac{1-b}{a}}$ as $\Psi$.
From the above relationship we can infer the critical relationship between a test's parameters and the angle $\beta$:

\begin{large}
\begin{equation}
\lim_{\Psi \to 0} \beta = 0
\end{equation}
\end{large}

We now have enough information to distinguish between the shapes of two different screening curves which have the same $\varepsilon$. Notably, the test whose angle $\beta$ is lower will take up a greater area in the screening plane and therefore will offer a greater positive predictive value for a given risk level. Otherwise stated: 

\begin{large}
\begin{equation}
\beta_2 < \beta_1\leftrightarrow \Psi_2 < \Psi_1 \rightarrow \rho_2 > \rho_1
\end{equation}
\end{large}

The latter follows from the non-commutative properties of $\varepsilon$. 

\subsection{Using the angle $\beta$ to determine the LR+ of a screening curve}

From the relationship above in (6), and knowing that the positive likelihood ratio (LR+) \cite{deeks2004diagnostic} is defined as the ratio of the sensitivity over the compliment of the specificity ($\frac{a}{1-b}$) we obtain:

\begin{large}
\begin{equation}
tan(\beta) = \frac{1}{\sqrt{\frac{a}{1-b}}} \Leftrightarrow \sqrt{\frac{a}{1-b}} = \frac{1}{tan(\beta)}=cot(\beta)
\end{equation}
\end{large}
\
\\
\
where $cot(\beta)$ is the cotangent of the angle $\beta$. Thus, a new geometric formulation for the positive likelihood ratio (LR+) \cite{deeks2004diagnostic} ensues, now defined as:

\begin{large}
\begin{equation}
cot^2{(\beta)} = \frac{a}{1-b}
\end{equation}
\end{large}

\section{Deriving the likelihood ratio LR+ by means of the prevalence threshold}

\begin{theorem}
Let S be the screening plane of area 1 with invariant points [0,0] and [1,1] where lies the screening curve's continuous function $0<\rho(\phi)<1$. Then there is a vertical function $f(a,b)$ which transects the screening curve at the prevalence threshold, such that the ratio $\chi$ between the derivatives $\frac{dy_1}{dx}$ and $\frac{dy_2}{dx}$ of the linear equations $y_1$, $y_2$ which stem from the invariant points on the plane to the prevalence threshold equals the positive likelihood ratio (LR+).
\end{theorem}

\subsection{Proof of Theorem 2}
While the angular definition makes intuitive sense, we can derive the positive likelihood ratio (LR+) by using the prevalence threshold as a transecting function that divides the screening curve into two halves. Notably, as stated in previous work defining the prevalence threshold, the half of the screening curve below the prevalence threshold represents the risk range where the reliability of the test in question drops exponentially and the number of false positive results increases most rapidly. This time, we will use the second invariant point, namely [1,1], to define the linear equation from the prevalence threshold [$\frac{\sqrt{1-b}}{\sqrt{a}+\sqrt{1-b}}$, $\sqrt{\frac{a}{1-b}}\frac{\sqrt{1-b}}{\sqrt{a}+\sqrt{1-b}}$] to this point. In so doing, we thus obtain:

\begin{large}
\begin{equation}
m = \frac{\Delta y}{\Delta x} = \frac{y_2-y_1}{x_2-x_1}=\frac{1-\sqrt{\frac{a}{1-b}}\frac{\sqrt{1-b}}{\sqrt{a}+\sqrt{1-b}}}{1-\frac{\sqrt{1-b}}{\sqrt{a}+\sqrt{1-b}}} 
\end{equation}
\end{large}
\
We can use the invariant point [1,1] to determine the y-intercept and thus obtain the following final linear equation for this line:
\begin{large}
\begin{equation}
y=\frac{\left(1-\frac{\sqrt{1-b}}{\sqrt{1-b}+\sqrt{a}}\sqrt{\frac{a}{1-b}}\right)}{1-\frac{\sqrt{1-b}}{\sqrt{1-b}+\sqrt{a}}}x\ +\ \left[1-\frac{\left(1-\frac{\sqrt{1-b}}{\sqrt{1-b}+\sqrt{a}}\sqrt{\frac{a}{1-b}}\right)}{1-\frac{\sqrt{1-b}}{\sqrt{1-b}+\sqrt{a}}}\right]
\end{equation}
\end{large}

If we take the ratio of the derivatives $\chi$ of both equations in (4) and (12) we obtain the following:

\begin{large}
\begin{equation}
\Rightarrow \chi =\frac{\frac{dy_1}{dx}}{\frac{dy_2}{dx}} = \frac{\sqrt{\frac{a}{1-b}}}{\frac{\left(1-\frac{\sqrt{1-b}}{\sqrt{1-b}+\sqrt{a}}\sqrt{\frac{a}{1-b}}\right)}{1-\frac{\sqrt{1-b}}{\sqrt{1-b}+\sqrt{a}}}} =\frac{a}{1-b}
\end{equation}
\end{large}

This is the equation for the positive likelihood ratio of a screening test (LR+). It thus follows that we can re-define the prevalence threshold as the single dividing point on the screening curve from which only two lines can de drawn to the invariant points on the screening curve such that the ensuing ratio of their derivative equals the positive likelihood ratio.  Graphically, we can depict the two linear functions in red and blue as follows:
\begin{center}

\hspace*{-8.4em}
\begin{tikzpicture}
 
	\begin{axis}[
    axis lines = left,
    xlabel = $\phi$,
	ylabel = {$\rho(\phi)$},
	   y label style={at={(current axis.above origin)},rotate=270,anchor=north},    
     ymin=0, ymax=1,
     xmin=0,xmax=1,
    legend pos = south east,
    width=10cm,
    height=10cm,
     ]
      \addplot [
	domain= 0.34:1,
	color= black,
	]
	{(0.95*x)/((0.95*x+(1-0.75)*(1-x))};
	 \addplot [
	domain= 0:0.34,
	color= black,
	]
	{(0.95*x)/((0.95*x+(1-0.75)*(1-x))};
	 \addplot [
	domain= 0:0.34,
	color= red,
	]
	{((0.95/(1-0.75))^0.5)*x};
	\addplot [
	domain= 0.34:1,
	color= blue,
	]
	{((0.339/(0.660))*x+0.489};
	 \addplot [
	domain= 0:0.34,
	color= red,
	style=dashed,
	]
	{0.66};
	 \addplot [
	domain= 0.34:1,
	color= blue,
	style=dashed,
	]
	{0.66};
	
	\addplot[samples=50, smooth,domain=0:6,red, name path=three, style=dashed] coordinates {(0.34,0)(0.34,0.66)};
	\addplot[samples=50, smooth,domain=0:6,blue, name path=three, style=dashed] coordinates {(1,0.66)(1,1)};
	\end{axis}
    \node[] at (47:4.6)  {$\phi_e$};
	\draw [decorate,decoration={brace,amplitude=10pt},xshift=-2pt,yshift=0pt]
(-1,0) -- (-1,5.5) node [red,midway,xshift=-2.8cm] 
{$\Delta y_a = \sqrt{\frac{a}{1-b}}\frac{\sqrt{a\left(-b+1\right)}+b-1}{(\varepsilon-1)}$};
\draw [decorate,decoration={brace,amplitude=10pt},xshift=-4pt,yshift=0pt]
(0.2,5.7) -- (3,5.7) node [red,midway,xshift=0.6cm,yshift=0.9cm] 
{\large $\Delta x_a = \frac{\sqrt{a\left(-b+1\right)}+b-1}{(\varepsilon-1)} $};
\draw [decorate,decoration={brace,amplitude=10pt}, xshift=275pt,yshift=7pt]
(-1,8) -- (-1,5.5) node [blue,midway,xshift=2.4cm, yshift = 0cm] 
{\tiny $\Delta y_b = 1-\sqrt{\frac{a}{1-b}}\frac{\sqrt{a\left(-b+1\right)}+b-1}{(\varepsilon-1)}$};
\draw [decorate,decoration={brace,amplitude=10pt},xshift=55pt,yshift=124.9pt]
(6.4,1) -- (1,1) node [blue,midway,xshift=0cm,yshift=-1.0cm] 
{\large $\Delta x_b = 1-\frac{\sqrt{a\left(-b+1\right)}+b-1}{(\varepsilon-1)} $};
\end{tikzpicture}
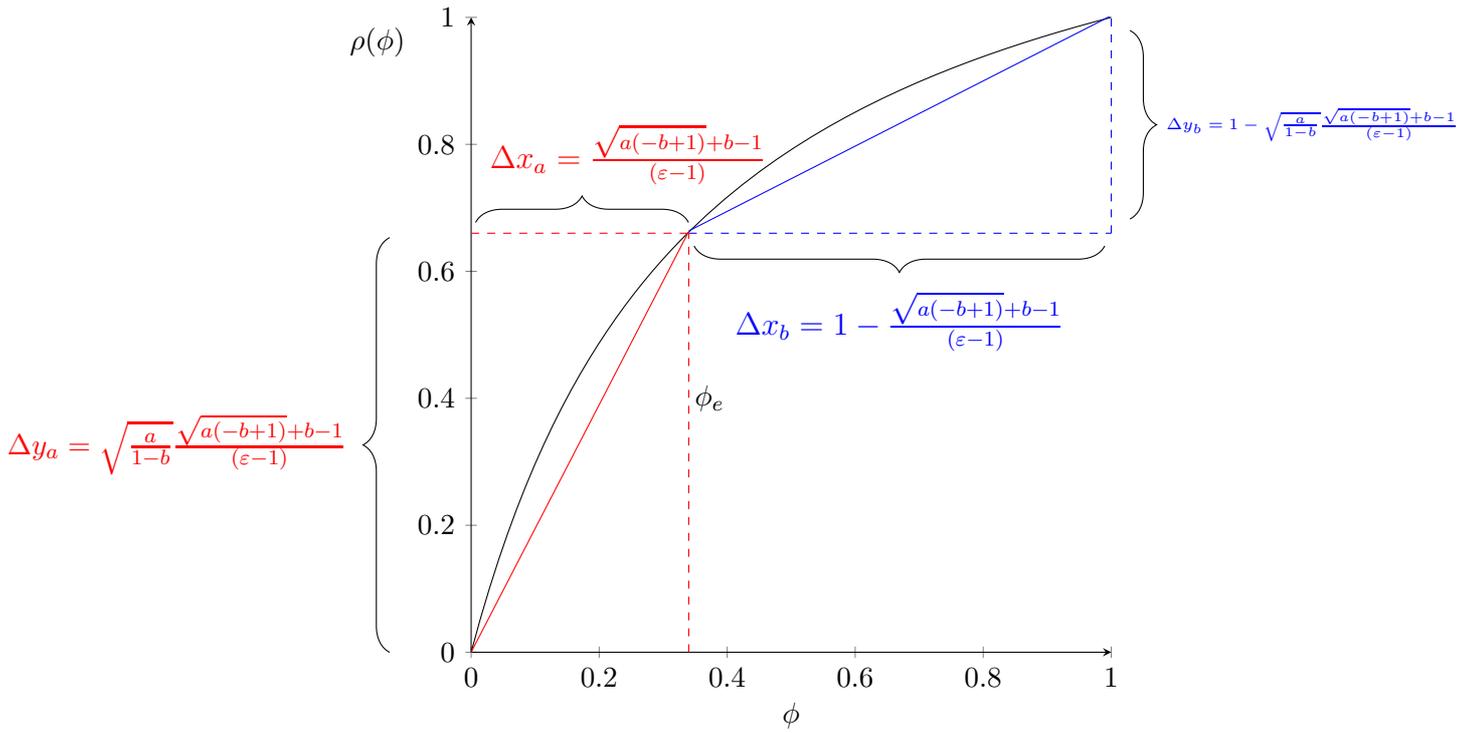
\textbf{Figure 6. Prevalence threshold transects the screening curve}
\\
\

\end{center}
\newpage
\section{Generalized version of the theorem on the Geometric Definition of the Positive Likelihood Ratio (LR+)}

\begin{theorem}
Let S be the screening plane of area 1 with invariant points [0,0] and [1,1] where lies the screening curve's continuous function $0<\rho(\phi)<1$. Then there is a vertical function $f(a,b)$ which transects the screening curve at any given point, such that the ratio $\chi$ between the derivatives $\frac{dy_1}{dx}$ and $\frac{dy_2}{dx}$ of the linear equations $y_1$, $y_2$ which stem from the invariant points on the plane to that point on the curve equals the positive likelihood ratio (LR+).
\end{theorem}

\subsection{Proof of Theorem 3}
While the use of the prevalence threshold as a transecting function that divides the screening curve into two halves can yield the LR+, this point is but a special case of a broader definition. Indeed, drawing a line from each invariant point to any point on the curve yields linear equations whose ratio of derivatives yield the LR+. This time, we will use the PPV equation to obtain the coordinates of any point on the curve as [$\phi,\frac{a\phi}{a\phi+\left(1-b\right)\left(1-\phi\right)}$]. Determining the equation of the lines that stem from the invariant origin [0,0] to any point on the curve we obtain:

\begin{large}
\begin{equation}
m_1 = \frac{\Delta y}{\Delta x} = \frac{\frac{a\phi}{ a\phi+(1-b)(1-\phi)}-0}{\phi-0} = \frac{\frac{a\phi}{ a\phi+(1-b)(1-\phi)}}{\phi}
\end{equation}
\end{large}
\
We can likewise use the invariant point [1,1] to determine the slope of the second line:
\begin{large}
\begin{equation}
m_2 = \frac{\Delta y}{\Delta x} = \frac{1-\frac{a\phi}{ a\phi+(1-b)(1-\phi)}}{1-\phi}
\end{equation}
\end{large}
If we take the ratio of the derivatives $\chi$ of both equations in (14) and (15) we obtain the following:

\begin{large}
\begin{equation}
\Rightarrow \chi =\frac{\frac{dy_1}{dx}}{\frac{dy_2}{dx}} = \frac{\frac{\frac{a\phi}{ a\phi+(1-b)(1-\phi)}}{\phi}}{\frac{1-\frac{a\phi}{ a\phi+(1-b)(1-\phi)}}{1-\phi}}=\frac{a}{1-b}
\end{equation}
\end{large}

Once again, we retrieve the equation for the positive likelihood ratio of a screening test (LR+). It thus follows that Theorem 3 generalizes the geometrization of the LR+ from the prevalence threshold point, as in Theorem 2, to any point on the screening curve.
\section{Discussion}
By having a foundational understanding of the interpretation of sensitivity, specificity, predictive values, and likelihood ratios, healthcare providers can better understand outputs from current and new diagnostic assessments, aiding in decision-making and ultimately improving healthcare for patients. That said, it is important to understand the limitations of the LR. First, and perhaps most obviously, the accuracy of a LR depends entirely upon the relevance and quality of the studies that generated the numbers (sensitivity and specificity) that inform that LR. Even when a LR is used in this fashion there are definite limits to the accuracy that we can presume underlies the number, for example the sensitivity and specificity evidence as originally generated may be flawed and the pre-test probability judgment can vary widely which mean that there are margins of error that should considered even under ideal circumstances. In addition, LRs have never been validated for use in series or in parallel. In other words there is no precedent to suggest that LRs can be used one after the other (i.e. using one LR to generate a post-test probability, and then using this as a pre-test probability for application of a different LR) or simultaneously, to arrive at a more accurate probability or diagnosis. It is important to keep these limitations in mind when using LRs because in many ways it is quite counter-intuitive to imagine that only one question at a time can be addressed when seeing a patient in the clinical environment with all of its inherent complexity. Despite this seemingly narrow use, LRs remain an invaluable and unique tool, as there is no other established method for adjusting a probability of disease based on known diagnostic test properties.
\section{Conclusion}
Using the concepts of the prevalence threshold and the invariant points on the screening plane, the work herein presented provides a geometric definition of the 
positive likelihood ratio (LR+) and describes a formal measure to compare the performance of two screening tests whose $\varepsilon$ are equal. Indeed, when faced with two screening tests whose sensitivity and specificity parameters add to the same value, the angle $\beta$ and the ratio of derivatives $\chi$ can help distinguish their performance throughout the prevalence/risk spectrum.

\newpage

\bibliographystyle{unsrt}
\bibliography{references}

\begin{thebibliography}{1}

\bibitem{manrai2014medicine}
Arjun~K Manrai, Gaurav Bhatia, Judith Strymish, Isaac~S Kohane, and Sachin~H
  Jain.
\newblock Medicine’s uncomfortable relationship with math: calculating
  positive predictive value.
\newblock {\em JAMA internal medicine}, 174(6):991--993, 2014.

\bibitem{balayla2020prevalence}
Jacques Balayla.
\newblock Prevalence threshold ($\phi$ e) and the geometry of screening curves.
\newblock {\em Plos one}, 15(10):e0240215, 2020.

\bibitem{birkhoff1922invariant}
George~David Birkhoff and Oliver~Dimon Kellogg.
\newblock Invariant points in function space.
\newblock {\em Transactions of the American Mathematical Society},
  23(1):96--115, 1922.

\bibitem{balayla2020derivation}
Jacques Balayla.
\newblock Derivation of generalized equations for the predictive value of
  sequential screening tests.
\newblock {\em arXiv preprint arXiv:2007.13046}, 2020.

\bibitem{hayden1999likelihood}
Stephen~R Hayden and Michael~D Brown.
\newblock Likelihood ratio: a powerful tool for incorporating the results of a
  diagnostic test into clinical decisionmaking.
\newblock {\em Annals of emergency medicine}, 33(5):575--580, 1999.

\bibitem{grove1984positive}
Daniel~M Grove.
\newblock Positive association in a two-way contingency table: likelihood ratio
  tests.
\newblock {\em Communications in Statistics-Theory and Methods},
  13(8):931--945, 1984.

\bibitem{deeks2004diagnostic}
Jonathan~J Deeks and Douglas~G Altman.
\newblock Diagnostic tests 4: likelihood ratios.
\newblock {\em Bmj}, 329(7458):168--169, 2004.

\end{thebibliography}

\end{document}